\documentclass[11pt]{article}
\usepackage{graphicx,amssymb}

\def\aap{Astron. Astrophys.}

\def\aj{Astron. J.}

\def\apj{Astrophys. J.}

\def\astl{Astron. Lett.}
\def\prl{Phys. Rev. Lett.}

\def\Nreal{\ensuremath{N_\mathrm{real}}}
\def\Nbg{\ensuremath{N_\mathrm{bg}}}

\def\Npsr{\ensuremath{N_\mathrm{PSR}}}
\def\Pbc{\ensuremath{P_\mathrm{bc}}}
\def\PN{\ensuremath{P_\mathrm{r}}}
\def\Ep{\ensuremath{E_\mathrm{p}}}
\def\EFe{\ensuremath{E_\mathrm{Fe}}}

\def\etal{{et~al.}}
\newcommand{\art}[5]{#1, #2 \textbf{#3}, #5 (#4).}
\newcommand{\artx}[6]{#1, #2 \textbf{#3}, #5 (#4); \texttt{arXiv:#6}.}
\newcommand{\prep}[3]{#1, \textit{#2}, \texttt{arXiv:#3}.}
\newcommand{\proc}[4]{#1, \textit{#2} (#3),        #4.}
\begin{document}
\begin{center}
{\Large\bfseries
A Search for Small-Scale Anisotropy\\ of PeV Cosmic Rays\\[2mm]
}
M. Yu.\ Zotov, G. V. Kulikov\\
Skobeltsyn Institute of Nuclear Physics,\\
Lomonosov Moscow State University,\\
Moscow 119234, Russia
\end{center}

\medskip
{\narrower
\noindent
Recent results of Milagro, Tibet, ARGO-YBJ and IceCube experiments
on the small-scale anisotropy of Galactic cosmic rays (CRs) with energies
from units up to a few hundred TeV arise a question on a possible
nature of the observed phenomenon, as well as on the anisotropy of
CRs at higher energies.
An analysis of a small-scale anisotropy of CRs with energies at around
PeV registered with the EAS MSU array presented in the article, reveals a
number of regions with an excessive flux.
A typical size of the regions varies from $3^\circ$ up to $12^\circ$.
We study correlation of these regions with positions of potential
astrophysical sources of CRs and discuss a possible origin of the
observed anisotropy.

}

\section*{Introduction}

In the last few years, a big interest was attracted to results of
Milagro (Atkins \etal, 2004; Saz Parkinson, 2005; Abdo \etal, 2007, 2008),
Tibet~AS$\gamma$ (Amenomori \etal, 2005, 2006, 2009),
Super-Kamiokande (Oyama, 2006; Guillian \etal, 2007) and
ARGO-YBJ (Di Sciascio, Iuppa, 2011) experiments on the anisotropy of
cosmic rays (CRs) with energies from units up to a few tens TeV at scales
up to $30^\circ$.
As is witnessed by the results published, a number of regions with either
an excess or deficit of the CR flux of the above energies is observed in
the Northern hemisphere, and the shape, location and the degree of
magnitude depend on the energy of CRs.
Recent observations held with IceCube demonstrate that the small-scale
anisotropy takes place in the Southern hemisphere either, and that
the observed sky map for cosmic rays with the median energy 20~TeV
differs significantly from that at 400~TeV (Abbasi \etal, 2011a,b).

The above results have attracted considerable attention because,
on the one hand, they shed additional light on the structure of
the interstellar magnetic field, and on the other hand, they provide
information to solve the problem of the origin of cosmic rays,
which remains open for many years.
Recall that the most widely spread model is the model of diffusive
acceleration of charged particles on shock fronts generated by
explosions of supernovae, see, e.g., the review by Hillas (2005).
It is thought that CRs can be accelerated in supernova remnants (SNRs)
up to energies of the order of 1~PeV and even higher energies under
certain circumstances (Ptuskin \etal, 2011).
Still, experimental data that prove that TeV--PeV CRs are accelerated in
SNRs are rather poor and are mostly confined to observations of SNRs
W44 and G120.1+1.4 (SN~1572, Tycho SNR) in gamma rays
(Giuliani \etal, 2011; Morlino, Caprioli, 2011).
Even in these cases, one can only speak about acceleration of hadrons
up to energies of about 0.5~PeV.
The situation with the majority of known Galactic SNRs remains unclear.
In any case, a proof of the fact that CRs are accelerated in supernova
remnants up to energies of the order of PeV will not imply that
these are supernovae that are the main or the only source of Galactic
CRs of these energies.
By this reason, other possible astrophysical sources of CRs but not
only isolated SNRs are considered.

Models most close to SNRs in the sense of particle acceleration
are those that employ OB-associations, i.e., clusters of the most
massive and hot stars of spectral classes~O and B (see the works by Bykov
\etal, (1995), Parizot \etal, (2004), Binns \etal, (2008)), and
superbubbles formed by them (see, e.g., Higdon and Lingenfelter, 2005).
There are some facts that witness in favour of CR acceleration in
superbubbles.
For example, recent observations performed with the orbital gamma-ray
telescope Fermi-LAT have demonstrated that a region located approximately
between OB-association Cyg~OB2 and SNR $\gamma$-Cygni is a place of
effective CR acceleration (Ackermann \etal, 2011).
On the other hand, data on the composition of CRs show that most likely,
superbubbles are not the main source of Galactic cosmic rays
(Prantzos, 2011).

Another possible source of CRs are pulsars. Shortly after pulsars were
discovered and identified with neutron stars, it was demonstrated
they can be effective accelerators of CRs (Gunn, Ostriker, 1969).
This has attracted interest immediately, and there appeared a number
of works that analysed a possible correlation between arrival
directions of CRs and coordinates of pulsars known at that time
(Andrews \etal, 1970; Suga \etal, 1971). Interest to pulsars as
possible sources of CRs with energies $\gtrsim10^{14}$~eV did not
vanish subsequently, see, e.g, works by Blasi \etal\ (2000), Giller
and Lipski (2002), Bhadra (2003, 2006), Erlykin and Wolfendale
(2004). Models considered included acceleration both in pulsars
surrounded by wind nebulae (e.g., the Crab nebula) and by isolated
pulsars. In particular, Bhadra (2003) considered the Geminga pulsar
as a possible candidate for being the single source of the knee at
3~PeV. Other authors argue that the presence of a pulsar is essential
for an effective acceleration of CRs in a supernova remnant (e.g.,
Neronov, Semikoz, 2012).

The aforementioned results on the small-scale anisotropy of TeV CRs
appeared to be rather unexpected because trajectories
of charged particles of cosmic rays of these energies are entangled
in interstellar magnetic fields so that their arrival directions to
Earth are expected to be isotropic.
Still, a comparison of arrival directions of CRs with location of their
possible astrophysical sources can help to reveal their origin
in case of certain configurations of magnetic fields as well as
in case when a considerable part of CRs consists of neutral particles.
There is also a point of view that the Galactic magnetic field can be
uniform at scales of the order $\gtrsim300$~pc, so that charged
particles can keep directions of their movement within this scale
(Oyama, 2006).
All this motivated us to perform a new analysis of the complete data
set of the EAS MSU array. In our previous works dedicated to
examination of arrival directions of CRs with energies in the PeV
range registered with the EAS MSU array (Zotov, Kulikov, 2009, 2010,
2011) and the EAS--1000 Prototype array (Kulikov, Zotov, 2004; Zotov,
Kulikov, 2004, 2007) we presented a whole number of small-scale
regions of excessive flux (REFs) of CRs.
In the present work, we confirm the existence of such regions in the
experimental data of the EAS MSU array with the statistical significance
from $3\sigma$ up to $4.95\sigma$ and analyse their correlation with
location of Galactic SNRs, pulsars, OB-associations and open clusters.
It happens that the majority of regions found can be matched to
astrophysical objects of at least one of the above types.
Possible nature of the observed anisotropy is discussed in conclusion.

\section{Experimental Data and Method of The Investigation}

The analysed data set consists of 513,602 showers registered with the
EAS MSU array in 1984--1990. A description of the array can be found
in the paper by Vernov \etal\ (1979).
All EAS selected for the analysis satisfy a number
of quality criteria and have zenith angles $\theta\le45^\circ$.
The median number of the full number of charged particles at the level
of observation equals approximately $1.6\times10^5$.
According to the modern models of hadron interactions and data
on mass composition of primary cosmic rays, this value corresponds to
the energy of primary protons $E_0\approx1.7\times10^{15}$~eV
with an accuracy of the order of 10\%--20\%.
An error in determination of arrival directions is estimated to be of
the order of $3^\circ$.

The investigation is based on the method by Alexandreas \etal\
(1991), which was developed for the analysis of arrival directions of
EAS registered with the CYGNUS array, and since then has been used
multiple times for the analysis of data of other experiments.
The idea of the method is as follows.
To every shower in the experimental data set, an arrival time
of another shower is assigned in a pseudo-random way.
After this, new equatorial coordinates $(\alpha,~\delta)$ are calculated
for the whole data set basing on the original zenith and azimuthal
angles $(\theta,\phi)$ and assigned arrival times. This results in
a ``mixed'' map of arrival directions that differs from the original one
but has the same distribution in declination~$\delta$.
The mixing of the real map is performed multiple times, and the mixed
maps are averaged in order to reduce the dependence of the
result on the choice of arrival times.
The method is based on an assumption that the resulting mean
``background'' map has most of the properties of an isotropic background,
and presents the distribution of arrival directions of cosmic rays that
would be registered with the array in case there is no anisotropy.
A measure of difference between any two regions of the two
maps located within the same boundaries is defined as
$S=(\Nreal-\Nbg)/\sqrt{\Nbg}$, where \Nreal\ and \Nbg\ are the number of
EAS inside the same region in the real and background maps
respectively.
As a rule, selection of regions of excessive flux is performed basing on
the condition $S>3$.

To search for REFs, we divided both maps into ``basic'' cells with the
size $0.5^\circ\times0.5^\circ$, which were used then to form larger
rectangular regions with the size $\Delta\delta=3^{\circ}\dots15^{\circ}$.
The width of these regions was chosen so that for a given $\Delta\delta$
and $\delta=0^{\circ}$ the area of the region differed from the area
of a square with the same height by at most 1/6, and was kept
approximately constant for growing~$\delta$.
A detailed description of the method can be found
in the paper by Zotov, Kulikov (2010).

The method of Alexandreas \etal\ does not provide a direct answer to the
question about the chance probability of appearance of an REF.
It can be calculated basing on the value of~$S$, which is an
estimate of the standard deviation of a sample and thus acts as a
statistical significance.
By this, it is implicitly assumed that the deviation of~\Nreal\ from~\Nbg\
has a Gaussian distribution.
Hence, it is assumed the chance probability of an REF to appear is less
than $1-0.9973$ providing it was selected at $S>3$.

One can also estimate the chance probability of an appearance of an REF
from the number of EAS inside, as is suggested by the following
simple method based on the binomial distribution (Zotov, Kulikov, 2009).
Let a shower axis getting inside a region be a success. The number of
trials equals the number of showers~$N$ in the given data set,
and an estimate of success for a fixed region equals
$\tilde p = \Nbg/N$, where~\Nbg\ is the number of showers in the region
of the background map. The assumption is based on the fact that in
the method of Alexandreas \etal, \Nbg\ is considered to be an
expected number of showers in a cell. Obviously, the chance
probability of finding exactly~\Nreal\ EAS in a region equals
\[
	P(\nu=\Nreal) = C(N,\Nreal) \tilde{p}^{\Nreal} (1 - \tilde{p})^{N-\Nreal},
\]
where $\nu$ is a random variable equal to the number of successes in the
binomial model, and $C(N,\Nreal)$ is the corresponding binomial
coefficient. It is more interesting to consider a probability that there
are at most~\Nreal\ showers in the region $\PN=P(\nu\le\Nreal)$. The
analysis performed and the data presented below demonstrate that values
of~\PN\ correlate well with the values of chance probabilities calculated
on the basis of the significance~$S$.

The regions of excessive flux of CRs found in the data set of the EAS MSU
array do not solely consist of basic cells with an excess of
registered EAS over the background flux with a significance $S>3$.
In a typical REF, there are basic cells with an excess as well as a deficit
of EAS with respect to the background level.
Employment of the pure criterion $S>3$ can result in a region selected
as an REF solely due to a huge irregularity of the distribution of
arrival directions in declination~$\delta$.
In order to improve the robustness of selection of REFs, we tried a
number of additional quantities.
The most productive of them is the probability
$\Pbc=P(\xi<N_\mathrm{bc}^+)$, calculated from the following binomial
model.
Let~$\xi$ be a random variable equal to the number of basic
cells of the given region with an excess of EAS over the background
values, and $N_\mathrm{bc}^+$ is the corresponding experimental value.
The number of trials equals the number of basic cells in the REF.
It is natural to assume the probability of success to be equal to~1/2.
In the results presented below, all REFs satisfy a condition
$\Pbc>0.9545$, which corresponds to the significance
$>2\sigma$ for the Gaussian distribution.
We thus reduced the chance that an REF is selected solely due to the
non-uniformity of the EAS distribution with respect to~$\delta$.

\section{The Main Results}

In what follows, we consider REFs found in the experimental data of
the EAS MSU array under the following conditions:
\[
	\Nreal>100, \quad S>3, \quad \Pbc>0.9545.
\]
These conditions are satisfied for 1073 cells with
$\Delta\delta=3^\circ\dots11.5^\circ$
that form 41 non-overlapping regions, see Fig.~1.
All the REFs are located in the strip with declination
$\delta=14.5^\circ\dots81^\circ$ and have an area in the range from
7.5 to 128 square degrees.
The number of EAS inside selected cells varies from 101 up to 5552, while
background values vary from 67 to 5326.
Some parameters of the REFs are presented in Table~\ref{table:REF}.
We remark that regions that consist of a single cell satisfying the
above conditions (REFs~3, 7, 27, 31, 34 and~36) are not ``outbursts''
but parts of larger regions with~$S$ slightly less than~3.

\begin{figure}[!ht]
 \centerline{\includegraphics[width=0.8\textwidth]{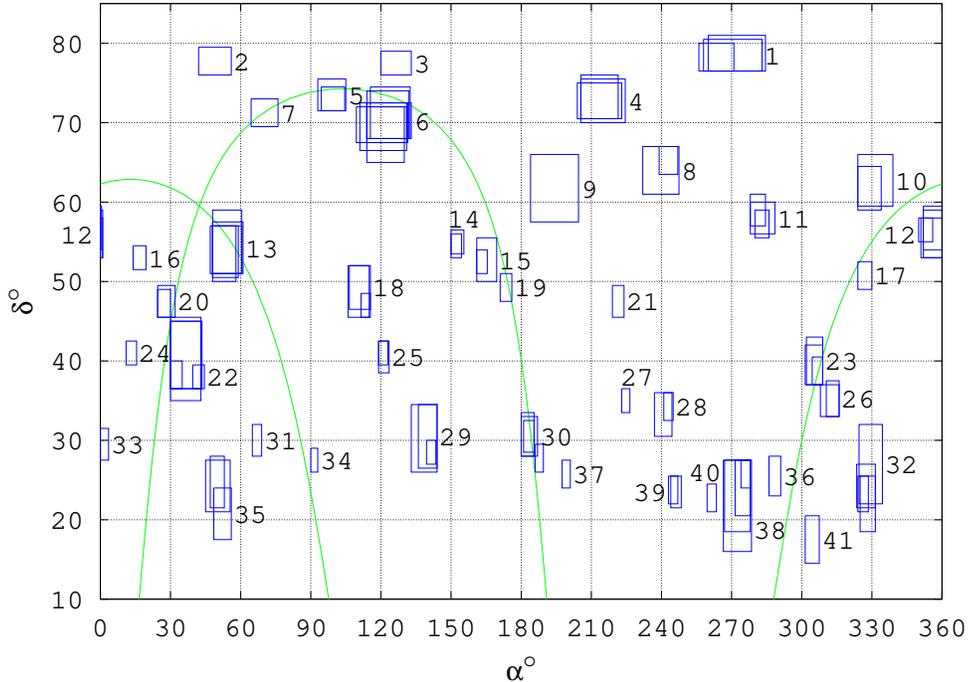}}
 \vspace{-7mm}
 \caption{%
 The regions of excessive flux of cosmic rays according to the data of the
 EAS MSU array, in equatorial coordinates.
 The curves at the left and right sides of the Figure show the Galactic plane,
 the $\cap$-like curve shows the Supergalactic plane.}
\end{figure}

It is interesting that regions 5, 6, 9, 13, 20, 21, 25, as well as
adjacent regions 23 and 26 are mainly made up of showers with the
particle number above the median value, while regions 8, 10, 12, 19,
22, 29, 32 and 38 with those below the median.

As is clear from Table~\ref{table:REF}, regions 40 and 41 are excluded
from the selection if one replaces the condition $\Nreal>100$ with
$\Nbg>100$.

\begin{table}[!ht]
\caption{Some parameters of the REFs of CRs according
to the data of the EAS MSU array.
Notation:
$\alpha$ and $\delta$ are the ranges of the corresponding equatorial
coordinates; \Nreal, \Nbg\ are the ranges of the experimental and
background flux of EAS in rectangular regions with excessive flux of
CRs that constitute the REF respectively; $\max S$, $\max\PN$,
$\max\Pbc$ are the maximum values of the respective variables for the
rectangular regions with $\Delta\delta\ge3^{\circ}$ that constitute
the REFs.}
\small
\begin{center}
\begin{tabular}{|c|c|c|c|c|c|c|c|}
\hline
REF & $\alpha,\,^\circ$ & $\delta,\,^\circ$ & \Nreal\ & \Nbg\ & $\max S$ & $\max\PN$ & $\max\Pbc$ \\
\hline
 1 & 256.0\dots284.5 & 76.5\dots81.0 &  442\dots1042 &  375.2\dots 942.8 & 3.93     & 0.999939 & 0.988350 \\
 2 &  42.0\dots 56.0 & 76.0\dots79.5 &  454\dots 598 &  393.3\dots 528.9 & 3.16     & 0.999081 & 0.995984 \\
 3 & 120.0\dots133.0 & 76.0\dots79.0 &  503          &  438.8            & 3.07     & 0.998763 & 0.990027 \\
 4 & 204.0\dots224.5 & 70.0\dots76.0 &  550\dots1852 &  480.2\dots1726.0 & 3.51     & 0.999741 & 0.996050 \\
 5 &  93.0\dots105.0 & 71.5\dots75.5 &  576\dots 861 &  507.0\dots 776.1 & 3.10     & 0.998903 & 0.996077 \\
 6 & 109.5\dots133.0 & 65.0\dots74.5 &  537\dots2727 &  470.0\dots2573.8 & 4.23     & 0.999985 & 0.999596 \\
 7 &  64.5\dots 76.0 & 69.5\dots73.0 &  853          &  768.3            & 3.06     & 0.998774 & 0.986488 \\
 8 & 232.0\dots247.5 & 61.0\dots67.0 &  654\dots2722 &  579.7\dots2565.9 & 3.47     & 0.999717 & 0.984226 \\
 9 & 184.0\dots204.5 & 57.5\dots66.0 & 4093\dots5552 & 3904.6\dots5326.2 & 3.09     & 0.999026 & 0.981344 \\
10 & 324.0\dots339.0 & 59.0\dots66.0 & 1418\dots2957 & 1309.0\dots2788.4 & 3.29     & 0.999453 & 0.993782 \\
11 & 278.0\dots288.5 & 55.5\dots61.0 &  518\dots1146 &  449.6\dots1025.1 & 4.48     & 0.999994 & 0.996891 \\
12 & 350.0\dots361.0 & 53.0\dots59.5 &  624\dots1820 &  549.6\dots1692.3 & 3.46     & 0.999684 & 0.989870 \\
13 &  47.0\dots 61.0 & 50.0\dots59.0 & 1607\dots3421 & 1481.8\dots3242.5 & 3.59     & 0.999817 & 0.991294 \\
14 & 150.0\dots155.5 & 53.0\dots56.5 &  511\dots 610 &  441.3\dots 539.5 & 3.32     & 0.999454 & 0.982209 \\
15 & 161.0\dots169.5 & 50.0\dots55.5 &  548\dots1464 &  476.8\dots1352.8 & 3.26     & 0.999344 & 0.995273 \\
16 &  14.0\dots 19.5 & 51.5\dots54.5 &  574\dots 627 &  504.7\dots 555.1 & 3.09     & 0.998853 & 0.974053 \\
17 & 324.0\dots330.0 & 49.0\dots52.5 &  735\dots 797 &  655.1\dots 714.6 & 3.12     & 0.998992 & 0.968514 \\
18 & 106.0\dots115.5 & 45.5\dots52.0 &  478\dots2200 &  415.8\dots2063.0 & 3.13     & 0.999061 & 0.993752 \\
19 & 171.0\dots176.0 & 47.5\dots51.0 &  647\dots 717 &  574.8\dots 638.8 & 3.09     & 0.998898 & 0.988769 \\
20 &  24.5\dots 32.0 & 45.5\dots49.5 &  479\dots 991 &  409.5\dots 897.6 & 3.88     & 0.999926 & 0.993358 \\
21 & 219.0\dots224.0 & 45.5\dots49.5 &  478\dots 767 &  416.2\dots 687.0 & 3.31     & 0.999439 & 0.997242 \\
22 &  30.0\dots 44.5 & 35.0\dots45.5 &  332\dots3905 &  279.0\dots3703.6 & 3.76     & 0.999884 & 0.985960 \\
23 & 301.5\dots309.0 & 37.0\dots43.0 &  425\dots1253 &  366.5\dots1148.0 & 3.28     & 0.999389 & 0.985377 \\
24 &  11.0\dots 15.5 & 39.5\dots42.5 &  373\dots 484 &  314.4\dots 404.1 & 3.97     & 0.999949 & 0.979888 \\
25 & 119.0\dots123.5 & 38.5\dots42.5 &  424\dots 618 &  366.2\dots 539.6 & 3.38     & 0.999564 & 0.999641 \\
26 & 308.0\dots316.0 & 33.0\dots37.5 &  294\dots 605 &  245.7\dots 533.9 & 3.63     & 0.999812 & 0.987770 \\
27 & 223.0\dots226.5 & 33.5\dots36.5 &  278          &  231.6            & 3.05     & 0.998636 & 0.978221 \\
28 & 237.0\dots245.0 & 30.5\dots36.0 &  256\dots 875 &  210.7\dots 789.1 & 3.35     & 0.999492 & 0.999657 \\
29 & 133.0\dots144.0 & 26.0\dots34.5 &  159\dots1613 &  121.6\dots1494.8 & 3.39     & 0.999508 & 0.994773 \\
30 & 180.0\dots189.5 & 26.0\dots33.5 &  162\dots 570 &  126.5\dots 501.8 & 3.55     & 0.999747 & 0.996845 \\
31 &  65.0\dots 69.0 & 28.0\dots32.0 &  292          &  241.4            & 3.26     & 0.999298 & 0.970029 \\
32 & 323.5\dots334.5 & 18.5\dots32.0 &  101\dots1148 &   67.3\dots1049.3 & 4.95     & 0.999998 & 0.999844 \\
33 &   0.0\dots  3.5 & 27.5\dots31.5 &  190\dots 221 &  150.6\dots 180.0 & 3.21     & 0.999141 & 0.990240 \\
34 &  90.0\dots 93.0 & 26.0\dots29.0 &  144          &  105.3            & 3.77     & 0.999858 & 0.967377 \\
35 &  45.0\dots 56.0 & 17.5\dots28.0 &  103\dots 481 &   75.4\dots 418.0 & 3.55     & 0.999716 & 0.999530 \\
36 & 286.0\dots291.0 & 23.0\dots28.0 &  264          &  218.1            & 3.11     & 0.998857 & 0.971556 \\
37 & 197.5\dots201.0 & 24.0\dots27.5 &  137\dots 155 &  103.6\dots 117.8 & 3.42     & 0.999551 & 0.957283 \\
38 & 266.5\dots278.5 & 16.0\dots27.5 &  144\dots 874 &  111.6\dots 789.3 & 3.20     & 0.999180 & 0.994773 \\
39 & 243.0\dots248.5 & 21.5\dots25.5 &  101\dots 160 &   72.9\dots 126.0 & 3.37     & 0.999451 & 0.990240 \\
40 & 259.5\dots263.5 & 21.0\dots24.5 &  106\dots 118 &   77.1\dots  88.1 & 3.30     & 0.999283 & 0.959287 \\
41 & 301.5\dots307.5 & 14.5\dots20.5 &  107\dots 131 &   78.6\dots  98.7 & 3.25     & 0.999199 & 0.977557 \\
\hline
\end{tabular}
\end{center}
\label{table:REF}
\end{table}

\subsection{Supernova Remnants}

As it was already discussed above, supernova remnants are the main
candidates for being the sources of Galactic CRs with energies
$\lesssim1$~PeV.
In Fig.~2, coordinates of Galactic SNRs located in the field of
observation are shown according to the catalogue by Green (2009).
Numbers denote REFs that have at least one SNR at the angular distance
$\le3^\circ$.

\begin{figure}[!ht]
 \centerline{\includegraphics[width=0.8\textwidth]{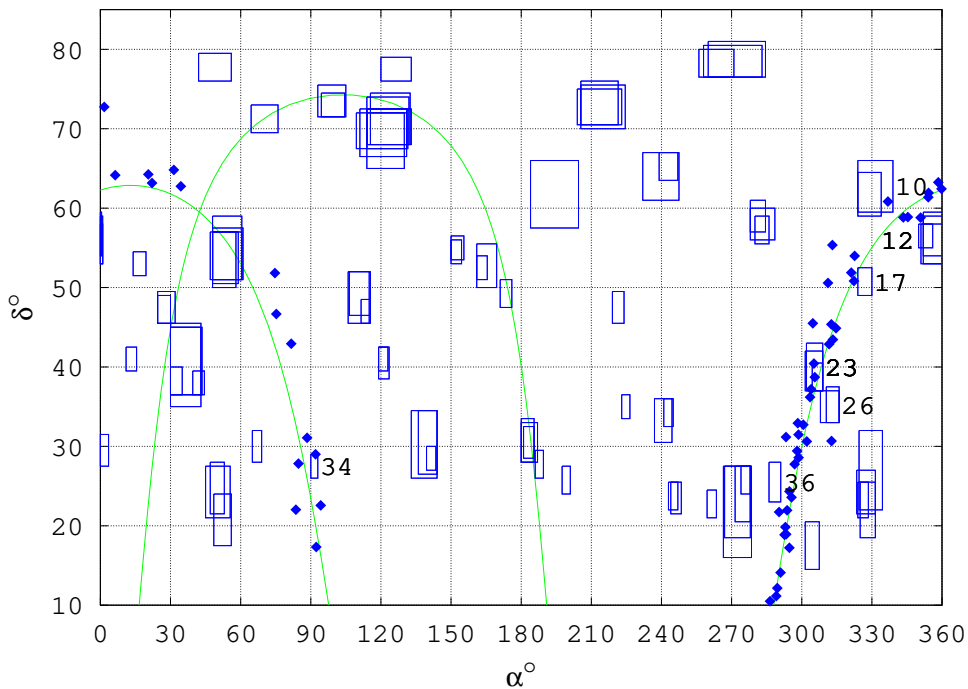}}
 \vspace{-7mm}
 \caption{%
 Coordinates of Galactic SNRs located in the field of
 observation, according to the catalogue by Green (2009).
 The numbers are shown for REFs that have at least one SNR in their
 3-degree neighbourhood.}
 \label{fig:SNR}
\end{figure}

There are seven such regions. All of them are located near the Galactic
plane. Twenty-one of 58 SNRs in the field of observation belong
to the 3-degree neighbourhoods of the regions of excessive flux.
Five SNRs lie inside REFs.
One of the SNRs (G76.9+1.0) is located in the vicinity of two REFs. It
lies inside REF 23 and, simultaneously, at the angular distance
$\rho\approx2.6^\circ$ from the adjacent REF 26.

Some parameters of the selected SNRs are presented in Table~\ref{table:SNR}.

\begin{table}[!ht]
\caption{%
Some parameters of SNRs in the 3-degree neighbourhood of the REFs.
Notation:
$\alpha,^\circ$, $\delta,^\circ$ are the equatorial coordinates of an
object;
$\rho,^\circ$ is the angular distance from an SNR to the given REF
(0 means the object is within the boundaries of the REF);
$d$ is the distance from the REF to the Solar system, kpc;
Size,$'$ is the angular size of the remnant, in arcminutes;
Type of the SNR: ``S'' or ``F'' if the remnant shows a ``shell'' or
``filled-centre'' structure, or ``C'' if it shows ``composite'' radio
structure with a combination of shell and filled-centre characteristics;
the question sign (?) after S, F, C denotes some uncertainty in the type
of an SNR; a single~? denotes objects that are regarded to be supernova
remnants though their nature is not well-understood (Green, 2009).}
\begin{center}
\begin{tabular}{|c|l|c|c|c|c|c|c|}
\hline
REF & Name        &$\alpha,^\circ$ & $\delta,^\circ$ & $\rho,^\circ$ & $d$, kpc  & Size,$'$ & Type\\
 &              &          &         &       &       &         &      \\
\hline
10 & G106.3+2.7   & 336.87   & 60.83   & 0     &       & 60x24   &  C?  \\
   & G108.2-0.6   & 343.42   & 58.83   & 2.36  & 3.2   & 70x54   &  S   \\
\hline
12 & G109.1-1.0 (CTB 109)
						& 345.40   & 58.88   & 2.56  & 3     & 28      &  S   \\
   & G111.7-2.1 (Cas A, 3C461)
                  & 350.86   & 58.80   & 0.59  & 3.4   & 5       &  S   \\
   & G113.0+0.2   & 354.15   & 61.37   & 1.87  & 3.1   & 40x17?  &  ?   \\
   & G114.3+0.3   & 354.25   & 61.92   & 2.42  & 0.7   & 90x55   &  S   \\
   & G116.9+0.2 (CTB 1)
                  & 359.79   & 62.43   & 2.93  & 1.6   & 34      &  S   \\
\hline
17 & G93.7-0.2 (CTB 104A)
                  & 322.33   & 50.83   & 1.05  & 1.5   & 80      &  S   \\
   & G94.0+1.0    & 321.21   & 51.88   & 1.72  & 5.2   & 30x25   &  S   \\
   & G96.0+2.0    & 322.62   & 53.98   & 1.69  & 4     & 26      &  S   \\
\hline
23 & G73.9+0.9    & 303.56   & 36.20   & 0.80  &       & 27      &  S?  \\
   & G74.9+1.2 (CTB 87)
                  & 304.01   & 37.20   & 0     &6.1--12& 8x6     &  F   \\
   & G76.9+1.0    & 305.58   & 38.72   & 0     &       & 9       &  ?   \\
   & G78.2+2.1 ($\gamma$-Cygni)
                  & 305.21   & 40.43   & 0     &       & 60      &  S   \\
   & G82.2+5.3 (CTB 88)
                  & 304.75   & 45.50   & 2.50  &       & 95x65   &  S   \\
   & G83.0-0.3    & 311.73   & 42.87   & 2.00  &       & 9x7     &  S   \\
\hline
26 & G74.0-8.5 (Cygnus Loop)
                  & 312.75   & 30.67   & 2.33  & 0.44  & 230x160 &  S   \\
   & G76.9+1.0    & 305.58   & 38.72   & 2.57  &       & 9       &  ?   \\
\hline
34 & G179.0+2.6   &  88.42   & 31.08   & 2.49  &       & 70      &  S?  \\
   & G182.4+4.3   &  92.04   & 29.00   & 0     &       & 50      &  S   \\
\hline
36 & G55.7+3.4    & 290.33   & 21.73   & 1.27  &       & 23      &  S   \\
   & G57.2+0.8    & 293.75   & 21.95   & 2.75  &       & 12?     &  S?  \\
\hline
\end{tabular}
\end{center}
\label{table:SNR}
\end{table}

The biggest number of SNRs is ``collected'' by REF 23, which is
located in the Galactic plane in the direction to a region in
constellation Cygnus with well-known X-ray binaries Cygnus X-1 and
Cygnus X-3, about 800 regions of ionized hydrogen HII, a big number
of Wolf--Rayet stars and a number of OB-associations.
Three supernova remnants lie inside this REF, three other are located
at angular distances from 0.8$^\circ$ to 2.5$^\circ$.
The most well-studied of them is $\gamma$-Cygni (G78.2+2.1), inside which
sources of X-ray and gamma-ray emission were registered.
A considerable attention is also attracted by SNR G76.9+1.0,
which contains a high-energy pulsar discovered recently
(Arzoumanian \etal, 2011), see below.
It is interesting that Cygnus X-3 lies within the boundaries of REF 23,
and a well-known X-ray source Cygnus X-1, which is likely to consist
of a massive black hole and a giant companion star, is only at
$2.4^\circ$ to the south, and is situated comparatively close to the Solar
system. The latest estimates give a distance of the order of 1.86~kpc
(Reid \etal, 2011).

REF 23 adjoins REF 26, which has another SNR located at
$\rho\approx2.3^\circ$ to the south, namely, Cygnus Loop (G74.0-8.5), that
is a subject of numerous investigations. Cygnus Loop contains a number
of X-ray and radio sources, and its distance to the Solar system is
estimated to be as small as 0.44~kpc.

The second in the number of neighbouring SNRs is region 12. The closest
remnant, located at only $\rho=0.6^\circ$, is Cassiopeia~A, one of
the most widely discussed candidates for being a source of Galactic
cosmic rays.
SNR G114.3+0.3, situated at 0.7~kpc from the Solar system, is also
worth mentioning.
It is interesting that all SNRs in the neighbourhood of this region but
Cassiopeia~A have pulsars within or close to their boundaries.
This is not true for the majority of SNRs located in the neighbourhood
of the other REFs.
Besides the SNRs in the vicinity of REF 12 and the aforementioned SNR
G76.9+1.0, only SNRs G106.3+2.7 (REF 10) and G55.7+3.4 (REF 36) are
associated with pulsars.

An extension of the analysed neighbourhood of the REFs up to $4^\circ$
leads to selecting seven more SNRs.
Among them, there is a well-known remnant IC443 (G189.1+3.0), which
contains an X-ray binary as a compact central object.
SNR IC443 is located at the angular distance $\rho\approx3.6^\circ$ to the
south from REF 34, and its distance from the Solar system is estimated
to be from 0.7~kpc to 2~kpc.
SNR S147, which is comparatively close to the Solar system, lies at
the angular distance $\rho\approx4.6^\circ$ to the west from REF 34.
The estimates of the distance to S147 vary from 0.36~kpc to 0.88~kpc.
It is interesting to mention that SNRs G106.3+2.7 (REF 10),
G109.1-1.0, G111.7-2.1 (Cas~A) (REF 12), G94.0+1.0 (REF 17),
G74.9+1.2, G78.2+2.1 ($\gamma$-Cygni) (REF 23), G74.0-8.5 (Cygnus
Loop) (REF 26) and IC443 have molecular clouds in their vicinity
(Jiang \etal, 2010).

\subsection{Pulsars}

\begin{figure}[!ht]
 \centerline{\includegraphics[width=0.8\textwidth]{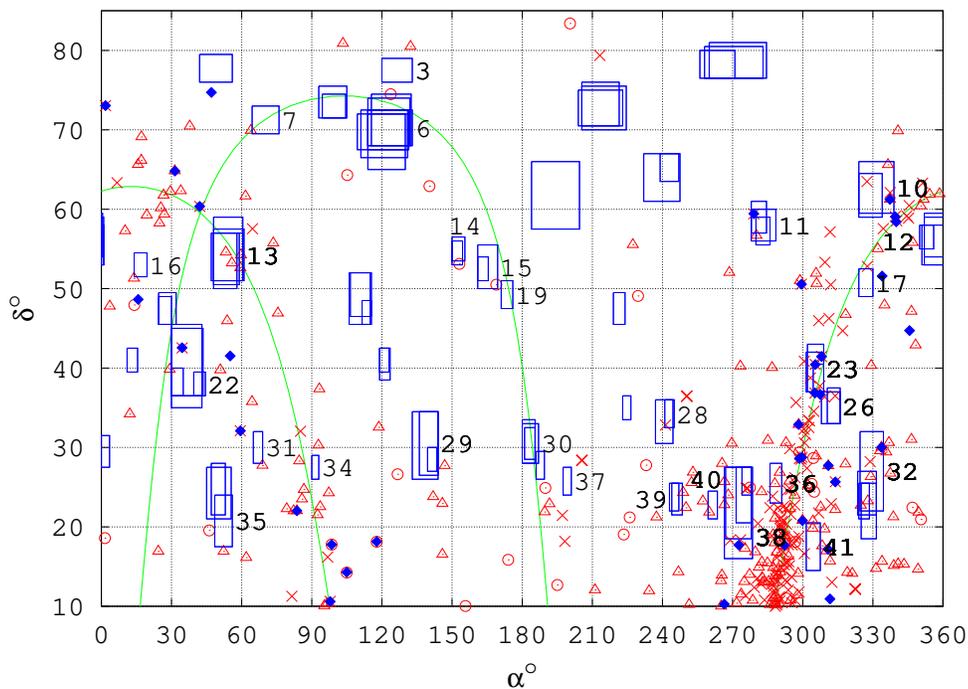}}
 \vspace{-7mm}
 \caption{%
 The REFs and Galactic pulsars (Manchester \etal, 2005).
 The numbers are shown for the REFs that have pulsars in their 3-degree
 neighbourhood.
 Different symbols are used for pulsars located at different distances~$d$
 from the Solar system:
 $\circ$ for $d\le1$~kpc, $\triangle$ for $1<d\le5$~kpc,
 $\times$ for pulsars at $d>5$ or at unknown distance.
 Dark diamonds denote coordinates of gamma-ray pulsars according to the
 Fermi LAT second source catalog (Abdo \etal, 2011).}
 \label{fig:PSRs}
\end{figure}

There were 355 Galactic pulsars known in the field $\delta>10^\circ$
at the time of the investigation (Manchester \etal, 2005), and 115 of
them were found to belong to the 3-degree neighbourhoods of 28 of the
REFs.
Fifty of them lie within the boundaries of 15 REFs, eight pulsars
are common for several regions, see Fig.~\ref{fig:PSRs}.
The number of selected pulsars grows up to 155 if one extends
the size of the studied neighbourhoods to $4^\circ$.
Remark that regions 6, 11 and 32 contain sub-regions selected with a
statistical significance $>4\sigma$, see our previous work (Zotov,
Kulikov, 2011).

Though pulsars are considered as possible sources of CRs since late
1960s, there is still no generally accepted model of acceleration of
charged particles in the vicinity of isolated pulsars.
To estimate the maximum energy of a particle accelerated near the light
cylinder of a pulsar, we employed an expression by Blasi \etal\ (2000)
that can be written as follows: $E_{\max}=0.34\,Z\,B\,\Omega^2$~eV,
where $Z$~is the charge number of the particle;
$B$~is the strength of the surface magnetic field:
$B = 3.2\times10^{19}\sqrt{P \dot P}$,~G;
$P$ is the barycentric period of rotation of the pulsar,~s;
$\Omega$ is the angular velocity of the pulsar, rad~s$^{-1}$,
and the radius of the pulsar equals $10^6$~cm.
The expression was obtained under the assumption that most of the
magnetic field energy in the pulsar wind zone is transformed into the
kinetic energy of the particles, and the density of electron-positron
pairs does not exceed $10^{-5}$ of that of ions of iron.

The maximum energy of protons accelerated in the selected 115 pulsars
varies from $2.2\times10^{12}$ to $1.0\times10^{16}$~eV.
For iron nuclei, the maximum energy varies from
$5.6\times10^{13}$ to $2.7\times10^{17}$~eV.
Parameters of some of the pulsars located in the 3-degree neighbourhoods of
the REFs are presented in Table~\ref{table:PSR}.
They include approximate values of the maximum energy of protons
and iron nuclei accelerated by these pulsars.
As can be seen from the data, in some cases (REFs 10, 22, 23, 36),
pulsars that lie in the vicinity of the REFs are able to accelerate
protons up to energies close to the knee.
Almost all such pulsars were registered not only in the radio but
also in the gamma-ray band.
In three of the above four cases, the regions of the excessive flux
are located near the Galactic plane.
Notice though that the majority of pulsars found in the vicinity of
the REFs are not able to accelerate protons up to energies of the
order of PeV.

The situation changes for iron nuclei.
For only nine of 28 REFs with pulsars in their 3-degree neighbourhood,
the respective pulsars are not able to accelerate iron nuclei up to
PeV energies (there are no data to make estimates
for pulsars in the vicinity of REFs 28, 29, 31 and~34).
Pulsars located in the vicinity of REFs
3, 6, 7, 15, 16, 19, 30, 35 and 37 (this is the same pulsar for REFs 15
and 19) have insufficient energy.
Pulsar B0053+47 (J0056+4756), which lies at 1~kpc from the Solar system,
is located at the angular distance $\rho\approx3.6^\circ$ from REF 16.
For this pulsar, $\max\EFe\approx2\times10^{15}$~eV.
Notice that all REFs with insufficiently energetic pulsars nearby
are located far from the Galactic plane.
The majority of them lie in the Supergalactic plane or its vicinity.

Let us mention a number of interesting pulsars located near the REFs.
Pulsar B2224+65 (J2225+6535, REF 10) is known to be associated with
the Guitar Nebula. The pulsar is moving fast relative to the
surrounding nebula and generates a bow shock.
Pulsars B2334+61 and J2021+4026 are associated with supernova remnants
G114.3+0.3 and G78.2+2.1 respectively, and young energetic pulsar
J2229+6114 (REF 10) is associated with SNR G106.3+2.7 and the
``Boomerang'' pulsar wind nebula (G106.6+2.9).
The pulsar and the nebula are sources of TeV gamma-ray emission
(Abdo \etal, 2009).
Pulsar B0355+54 (REF 13) is also associated with a nebula.
Note that acceleration of particles can involve other processes than
acceleration on the light cone if a pulsar is surrounded by a nebula.
In this case, processes typical for supernova remnants can play their
role.

We also remark the recently discovered pulsar J2022+3842 (Arzoumanian
\etal, 2011), not included in the ATNF catalogue yet. The pulsar lies
inside REF 23. It is located in supernova remnant G76.9+1.0 and is
observed both in radio and X-ray ranges.
It is the second after the pulsar in the Crab nebula
most energetic pulsar known in the Galaxy.
The ATNF catalogue does not yet contain gamma-ray pulsar J0308+7442 from
the Fermi LAT second source catalog (Abdo \etal, 2011) either.
The pulsar is located at the angular distance $\approx1^\circ$ from REF 2,
see Fig.~\ref{fig:PSRs}.

\begin{table}[!ht]
\caption{Some parameters of the pulsars located in 3-degree
neighbourhoods of the REFs (Manchester \etal, 2005).
Notation:
\Npsr{} is the number of pulsars in the neighbourhood of the given REF;
$\rho,^\circ$ is the angular distance from a pulsar to the REF;
$d$ is the distance from a pulsar to the Solar system,~kpc;
$P$ is the barycentric period of rotation of a pulsar,~s;
$\max\Ep$ and $\max\EFe$ are the maximum energies of a proton and
an iron nucleus accelerated near the light cylinder of a pulsar,~eV.
In the ``Notes'' column:
$\gamma$ denotes a pulsar with gamma-ray emission,
X~is an isolated neutron star with X-ray emission but no detectable
radio emission,
HE is a pulsar with pulsed emission from radio to infrared or higher
energies,
NR is a pulsar with pulsed emission only at infrared or higher energies.}
\small
\begin{center}
\begin{tabular}{|c|c|l|c|c|c|c|c|c|c|}
\hline
REF    & \Npsr{} & Name    &$\rho,^\circ$&$d$,&$P$, s& Age,               &   $\max\Ep$,         &   $\max\EFe$,        &  Notes\\
       &    &              &         & kpc &        &   years             &   eV                 &   eV                 &       \\
\hline
3      & 2  & B0841+80     &  1.48   & 3.38&   1.602&    5.69$\times 10^7$ &   4.48$\times10^{12}$ &   1.16$\times10^{14}$ &       \\
\hline
6      & 1  & B0809+74     &  0      & 0.43&   1.292&    1.22$\times 10^8$ &   3.79$\times10^{12}$ &   9.86$\times10^{13}$ &       \\
\hline
7      & 1  & B0410+69     &  0.18   & 1.57&   0.391&    8.08$\times 10^7$ &   1.54$\times10^{13}$ &   4.00$\times10^{14}$ &       \\
\hline
10     & 8  & B2224+65     &  0      & 2.00&   0.683&    1.12$\times 10^6$ &   7.49$\times10^{13}$ &   1.95$\times10^{15}$ &     \\
       &    & J2229+6114   &  0      & 0.8--3.0&   0.052&    1.05$\times 10^4$ &   1.02$\times10^{16}$ &   2.66$\times10^{17}$ &  $\gamma$,HE \\
       &    & J2238+59     &  0.51   & --- &   0.163&    2.62$\times 10^4$ &   2.05$\times10^{15}$ &   5.34$\times10^{16}$ &  $\gamma$,NR\\
\hline
11     & 2  & J1836+5925   &  0      & --- &   0.173&    1.84$\times 10^6$ &   2.30$\times10^{14}$ &   5.99$\times10^{15}$ &  $\gamma$,X,NR \\
\hline
12     & 6  & B2334+61     &  2.35   & 2.47&   0.495&    4.09$\times 10^4$ &   5.39$\times10^{14}$ &   1.40$\times10^{16}$ &       \\
       &    & B2319+60     &  1.19   & 3.21&   2.256&    5.08$\times 10^6$ &   1.06$\times10^{13}$ &   2.76$\times10^{14}$ &       \\
\hline
13     & 5  & B0355+54     &  0      & 1.10&   0.156&    5.64$\times 10^5$ &   4.60$\times10^{14}$ &   1.20$\times10^{16}$ &  HE \\
\hline
14     & 1  & J1012+5307   &  0      & 0.52&   0.005&    4.86$\times 10^9$ &   1.48$\times10^{14}$ &   3.84$\times10^{15}$ &  HE   \\
\hline
15     & 1  & B1112+50     &  0      & 0.54&   1.656&    1.05$\times 10^7$ &   1.01$\times10^{13}$ &   2.62$\times10^{14}$ &       \\
\hline
16     & 1  & B0052+51     &  0.21   & 2.40&   2.115&    3.51$\times 10^6$ &   1.37$\times10^{13}$ &   3.55$\times10^{14}$ &       \\
\hline
17     & 2  & B2148+52     &  0.30   & 5.67&   0.332&    5.21$\times 10^5$ &   2.25$\times10^{14}$ &   5.85$\times10^{15}$ &       \\
\hline
22     & 2  & J0218+4232   &  0      & 5.85&   0.002&    4.76$\times 10^8$ &   1.07$\times10^{15}$ &   2.77$\times10^{16}$ &  $\gamma$,HE \\
\hline
23     & 9  & J2032+4127   &  0      & 5.15&   0.143&    1.13$\times 10^5$ &   1.13$\times10^{15}$ &   2.93$\times10^{16}$ &  $\gamma$,HE \\
       &    & J2021+4026   &  0      & --- &   0.265&    7.68$\times 10^4$ &   7.36$\times10^{14}$ &   1.91$\times10^{16}$ &  $\gamma$,NR\\
       &    & J2021+3651   &  0.15   &18.88&   0.104&    1.72$\times 10^4$ &   3.98$\times10^{15}$ &   1.03$\times10^{17}$ &  $\gamma$,HE \\
\hline
26     & 5  & B2053+36     &  0      & 5.88&   0.222&    9.51$\times 10^6$ &   7.91$\times10^{13}$ &   2.06$\times10^{15}$ &       \\
\hline
28     & 1  & J1605+3249   &  0      & --- &   6.880&    ---              &   ---                &   ---                &  X,NR \\
\hline
29     & 2  & J0927+23     &  2.22   & 2.52&   0.762&    ---              &   ---                &   ---                &       \\
\hline
30     & 1  & B1237+25     &  1.17   & 0.86&   1.382&    2.28$\times 10^7$ &   8.22$\times10^{12}$ &   2.14$\times10^{14}$ &       \\
\hline
31     & 1  & J0435+27     &  0.27   & 3.43&   0.326&    ---              &   ---                &   ---                &       \\
\hline
32     & 8  & J2229+2643   &  2.65   & 1.43&   0.003&   3.23$\times10^{10}$&   1.01$\times10^{14}$ &   2.62$\times10^{15}$ &       \\
       &    & J2151+2315   &  0      & 1.42&   0.594&    1.33$\times 10^7$ &   2.50$\times10^{13}$ &   6.50$\times10^{14}$ &       \\
\hline
34     & 1  & J0611+30     &  1.27   & 3.55&   1.412&    ---              &   ---                &   ---                &       \\
\hline
35     & 2  & B0301+19     &  1.45   & 0.95&   1.388&    1.70$\times 10^7$ &   9.48$\times10^{12}$ &   2.47$\times10^{14}$ &       \\
\hline
36     &16  & B1935+25     &  2.93   & 2.76&   0.201&    4.95$\times 10^6$ &   1.21$\times10^{14}$ &   3.14$\times10^{15}$ &       \\
       &    & J1912+2525   &  0      & 2.02&   0.622&    4.38$\times 10^7$ &   1.32$\times10^{13}$ &   3.42$\times10^{14}$ &       \\
       &    & B1930+22     &  2.04   & 9.80&   0.144&    3.98$\times 10^4$ &   1.88$\times10^{15}$ &   4.88$\times10^{16}$ &       \\
\hline
37     & 1  & J1308+2127   &  2.56   & --- &  10.313&    1.46$\times 10^6$ &   4.34$\times10^{12}$ &   1.13$\times10^{14}$ &  X,NR    \\
\hline
38     &14  & J1829+2456   &  0      & 0.75&   0.041&    1.24$\times10^{10}$&  1.18$\times10^{13}$ &   3.07$\times10^{14}$ &       \\
       &    & J1752+2359   &  0      & 2.70&   0.409&    1.01$\times 10^7$ &   4.16$\times10^{13}$ &   1.08$\times10^{15}$ &       \\
\hline
39     & 2  & B1633+24     &  0.33   & 2.27&   0.491&    6.51$\times 10^7$ &   1.37$\times10^{13}$ &   3.55$\times10^{14}$ &       \\
       &    & J1640+2224   &  1.45   & 1.19&   0.003&    1.77$\times10^{10}$&  1.28$\times10^{14}$ &   3.34$\times10^{15}$ &       \\
\hline
40     & 4  & J1709+2313   &  2.05   & 1.83&   0.005&    2.02$\times10^{10}$&  8.20$\times10^{13}$ &   2.13$\times10^{15}$ &       \\
       &    & J1720+2150   &  0      & 3.59&   1.616&    3.46$\times 10^7$ &   5.71$\times10^{12}$ &   1.48$\times10^{14}$ &       \\
\hline
41     &10  & B1957+20     &  1.52   & 1.53&   0.002&    1.51$\times 10^9$ &   8.68$\times10^{14}$ &   2.26$\times10^{16}$ &  $\gamma$,HE \\
\hline
\end{tabular}
\end{center}
\label{table:PSR}
\end{table}
\clearpage

Three more regions (4, 20 and 21) obtain nearby pulsars if the size of
the considered neighbourhoods of the REFs is extended up to $4^\circ$.
Pulsar J1412+7922 (Calvera), the nature of which is not completely
clear (Halpern, 2011), lies at $3.4^\circ$ from region~4.
The data available allow one to obtain the following estimate:
$\max\EFe\approx5\times10^{14}$~eV.
Gamma-ray pulsar J0218+4232, located inside REF 22, becomes a
neighbour of REF 20 with the angular distance $\rho=3.5^\circ$.
Finally, region 21 ``obtains'' pulsar J1518+4904, situated at only
0.7~kpc from the Solar system.
It is necessary to note though that this is an old pulsar with the
age estimated to be $2.4\times10^{10}$ years.
For it, $\max\EFe\approx2\times10^{14}$~eV.


\subsection{Open Clusters, OB-associations}

\begin{figure}[!ht]
 \centerline{\includegraphics[width=0.8\textwidth]{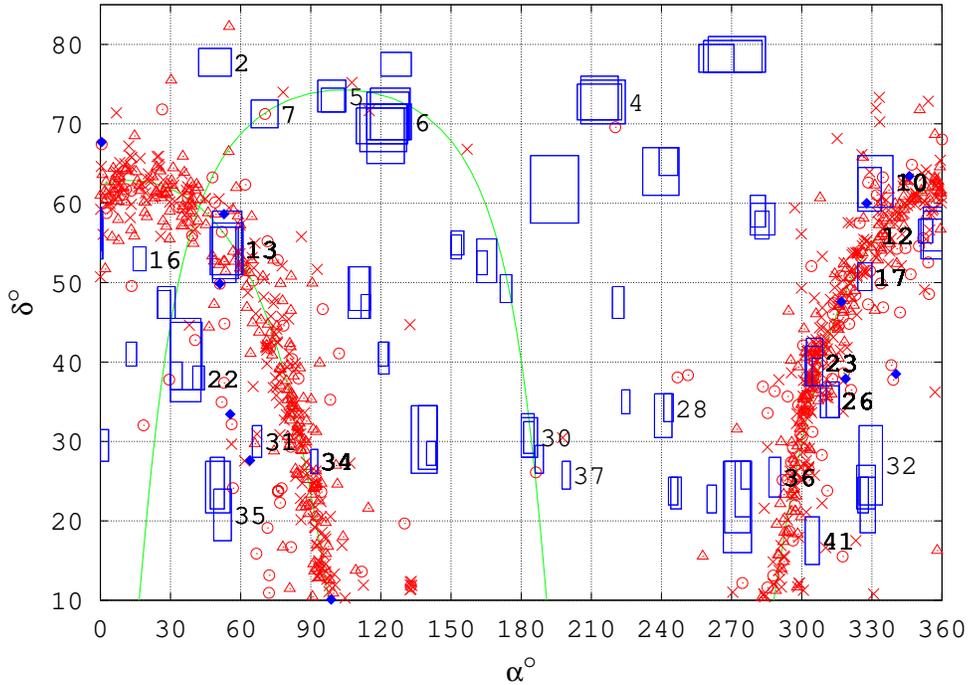}}
 \vspace{-7mm}
 \caption{%
 Galactic open clusters and nearby OB-associations.
 The numbers are shown for the REFs that have open clusters in their
 3-degree neighbourhood.
 Different symbols are used for open clusters located at different
 distances~$d$ from the Solar system:
 $\circ$ for $d\le1$~kpc, $\triangle$ for $1<d\le5$~kpc,
 $\times$ for open clusters at $d>5$ or at unknown distance.
 Dark diamonds show coordinates of OB-associations out to a
 distance of 1~kpc from the Solar system.}
 \label{fig:OpC}
\end{figure}

The location of Galactic open clusters (Dias \etal, 2002)
and OB-associations out to a distance of 1~kpc from the Solar system
(de~Zeeuw \etal, 1999) is shown in Fig.~\ref{fig:OpC}.
There are 343 open clusters in the 3-degree neighbourhood of the
REFs with 23 of them being common for two regions.
As is clear from the Figure, the majority of the open clusters lie in
the Galactic plane and due to this get in the vicinity of all the REFs
located in this part of the celestial sphere.
A comparison with Fig.~\ref{fig:SNR} and Fig.~\ref{fig:PSRs} reveals
though that all these REFs have not only open clusters nearby but also
Galactic pulsars and part of them have SNRs.
Still, let us again attract attention to REF 23 and its companion REF
26, which are located in the direction to an active star-forming
region in constellation Cygnus that contains a big number of
OB-associations, Cyg OB2 in particular.
This is the part of the sky where observations from the Fermi LAT
orbital gamma-ray telescope have recently revealed an extended cocoon
of CRs freshly accelerated up to energies 80--300~TeV for protons
(Ackermann \etal, 2011).

Notice a number of interesting coincidences.
REF 2, which has neither pulsars nor supernova remnants in its vicinity,
has two open clusters nearby, namely, Berkeley~8 and SAI~30.
They lie far away from the Solar system at the distance $d\approx3.1$~kpc.
Open cluster Collinder 285, which is only out at the distance of 25~pc
from the Solar system, is located practically at the southern boundary
of REF 4 ($\rho=0.4^\circ$).
REF 5 also obtains a neighbour with open cluster OCl 374 located
at the angular distance $0.65^\circ$.

Three REFs have nearby open clusters that are located close to
the Solar system.
These are Melotte 111 ($d=96$~pc, REF 30), M45 (Pleiades, $d=133$~pc,
REF 35) and Melotte 20 ($d=185$~pc, REF 13).
Open cluster Platais~3, which lies at the distance of 200~pc from the Solar
system, gets almost in the center of REF 7.
On the whole, there are open clusters located at distances less than 1~kpc
from the Solar system inside or in the close vicinity of 15 regions.
Among them, notice a well-known cluster Berkeley~87, located in the
Cygnus arm at a distance of 633~pc from the Solar system.

In the vicinity of the REFs, there are a number of OB-associations
located close to the Solar system.
We mention Cas-Tau association at the distance of 140~pc (REF 31)
and $\alpha$ Persei (Per OB3) at the distance of 170~pc (de~Zeeuw
\etal, 1999).

Remark finally that there are no known astrophysical objects of the
types considered above in the neighbourhood of REFs
1, 8, 9, 18, 24, 25, 27 and~33.

\section{Discussion and Conclusions}

We begin the discussion of the presented results with a comparison with
results of other experiments.
After the first analysis of the data of the EAS MSU array, we pointed out
that REFs 4, 5, 6, 10, 11, 12, 16, 17, 22, 30 and~33 practically coincide
or have intersecting boundaries with the regions found in the data of
the EAS--1000 Prototype array (Zotov, Kulikov, 2009).
A small overlapping of the boundaries also had place for REF~8.
All these results are confirmed.

There are a number of coincidences with regions of excessive flux of CRs,
found in the data obtained with the ``Klara-Chronotron'' station located
in Tian-Shan and aimed to study anisotropy of the cosmic rays at
energies $10^{14}$--$10^{15}$~eV (Nesterova \etal, 2011).
In particular, four regions were found with the statistical significance
$>3\sigma$ for a data set that corresponds to the primary energy of
protons at around $6\times10^{14}$~eV.
Two of them partially overlap with REFs 6 and 8.
Two other are located just near REFs 22 and 39.
It is surprising that in the case of REF 8 the overlapping is observed
for a region that does not have known possible Galactic sources of
PeV CRs nearby.
There are also a few coincidences with regions selected with the
statistical significance (2.1--3)$\sigma$.
REFs 13, 14, 15, 17, 28, 30 and 34 can be listed among them.

As it was already discussed above, a special interest to a search of
a small-scale anisotropy of PeV cosmic rays is caused by the recent
discoveries of the anisotropy of gamma-rays and CRs with energies
in the range from a few units up to several hundred TeV obtained
by the Milagro, Tibet AS$\gamma$ and ARGO-YBJ collaborations.
It is remarkable that five of eight gamma-ray point sources registered
with the Milagro experiment with the statistical significance $>4\sigma$
and located in the part of the celestial sphere considered in the present
article (Atkins \etal, 2004) are situated in the vicinity of the REFs.
Both REFs 23 and 35 have a source inside, one source lies at the
angular distance $0.7^\circ$ to the south from REF 26, two sources
are located at the angular distance $\approx3^\circ$ from (empty!)
REF~33.
An extended region of TeV gamma-ray emission found later in the Cygnus
arm with the statistical significance of $5.5\sigma$ (Saz~Parkinson, 2005)
also lies inside REF~23.
The second after the Crab pulsar brightest Galactic source of TeV
gamma-ray emission MGRO J2019+37 (Abdo \etal, 2007) is located at the
boundary of REF~23.

In addition to the extended region in the Cygnus arm, two other large
regions of excessive flux (``Region~A'' and ``Region~B'') dubbed as
``Milagro hot spots'' were found in the Milagro experiment (Abdo
\etal, 2007, 2008).
The regions have an angular size $\gtrsim10^\circ$ and are selected with
the statistical significance greater than $12\sigma$.
An analysis performed by the Milagro collaboration demonstrated that
the source of these regions is an excessive flux of hadrons with
energies that correspond to the energy of 10~TeV of a proton.
One of the regions is located to the south-west from REF 35.
Another one occupies a region approximately between REFs 25 and~29.
Similar regions in the energy range from a few TeV up to a few dozen
TeV were found in the Tibet AS$\gamma$ (Amenomori \etal, 2005, 2006,
2009) and ARGO-YBJ (Di~Sciascio, Iuppa, 2011) experiments.
In particular, the ARGO-YBJ results show an excessive flux of CRs
near REFs 4, 22 and~25.
It is interesting that one of the regions of excessive flux of
TeV gamma-ray emission registered with the Tibet AS$\gamma$ array
with the statistical significance $>4.5\sigma$ is located near empty REF~27.

There are no extended regions of excessive flux in the Tibet AS$\gamma$
data at energies of the order of 300~TeV pronounced as clearly as at
lower energies (Amenomori \etal, 2006).
Still, an analysis of the presented intensity maps reveals a few
interesting details.
In particular, it happens there is an extended region near the
Supergalactic plane around $\alpha\sim120^\circ$, $\delta\sim70^\circ$,
i.e., at the place of location of REF 6, which includes subregions
with an excess $>4\sigma$.

Up to now, there is no satisfactory model that could explain the origin
of the small-scale anisotropy of TeV CRs.
The Milagro collaboration and a number of other authors tend to a point
of view that the regions found appeared due to a special configuration
of magnetic fields and related effects in the heliosphere and at its
boundary, or at the adjacent region of the Galaxy at distances within
a few dozen parsecs
(Drury, Aharonian, 2008; Battaner \etal, 2009; Malkov \etal, 2010;
Amenomori \etal, 2010; Giacinti, Sigl, 2011).
It is also possible that the anisotropy is caused by the distribution
of CRs sources or by propagation effects through the regions of local
turbulence generated by the interaction of heliospheric and
interstellar magnetic fields (Desiati, Lazarian, 2011).
A question whether the models suggested in these works can be extended
from TeV energies up to the knee remains open.

If one assumes there are Galactic magnetic fields of the necessary
configuration, then there arises a question of their own origin.
Another question is why regions of excessive flux are found in the
vicinity of some of the possible sources of PeV CRs and are not
near the other.
Is this due to the properties of these astrophysical objects and the
surrounding media or to the properties of the magnetic fields?
The questions stated need a special consideration.
Since the distances to the pulsars given in Table~\ref{table:PSR} are
much greater than the mean free path of neutrons of PeV energies, and
the amount of electro-magnetic showers is not enough to form REFs,
there appears a question of how to keep the direction of charged CRs
moving through the Galactic magnetic field.
One of the possible explanations possibly relates to magnetic lenses.

A classical question of the cosmic ray physics, besides
the question of their origin, is their mass composition.
In connection with the regions found and the fact that the majority of
pulsars in their vicinity can only accelerate sufficiently heavy
nuclei to PeV energies, it is interesting to note that the analysis of
the Tibet AS$\gamma$ data lead to a conclusion that these are not
protons that dominate in the region of the knee but nuclei heavier
than helium (Amenomori, 2006b, 2008).
A similar conclusion was made basing on the data obtained at
mount Chacaltaya (Tokuno \etal, 2008).

We believe the results of the present research together with our
earlier works allow us to make a conclusion that there exists a
small-scale anisotropy of cosmic rays with energies at around PeV
and a typical angular size from $3^\circ$ up to $12^\circ$.
Location of a considerable number of the regions of excessive flux on
the celestial sphere correlates with coordinates of the Galactic SNRs,
pulsars, OB-associations and open clusters.
In our opinion, the correlation of REFs with multiple potential
sources of PeV CRs can witness that Galactic cosmic rays registered
on Earth were not born in a single source but in many
astrophysical sources of different nature.
Questions of how to preserve a direction of movement of CRs from their
sources to Earth in case these astrophysical objects give birth
to the REFs, and how to explain the regions without any known possible
sources of CRs in their neighbourhood, remain open and their solution
probably needs more detailed information on the structure of the
interstellar magnetic field in the vicinity of the Solar system.

\medskip
Only free, open source software was used for the investigation. In
particular, all calculations were performed with GNU Octave
running in Linux. The research has made use of the SIMBAD
database, operated at CDS, Strasbourg, France
(\texttt{http://simbad.u-strasbg.fr/simbad}).

The research was partially supported by the Russian Foundation for
Fundamental Research grant No.~11-02-00544{} and Federal contract
No.~16.518.11.7051.


\end{document}